\documentclass[12pt,preprint]{aastex}
\usepackage{emulateapj5}
\usepackage{epsfig}
\usepackage{apjfonts}

\makeatletter

\newenvironment{figurehere}
  {\def\@captype{figure}}
  {}
\makeatother

\newcommand{\mdot}{M$_{\odot}$}
\newcommand{\mv}{$M_V$}

\newcommand{\xopt}{$\mathrm{X5_{opt}}$}
\newcommand{\uopt}{$\mathrm{U_{opt}}$}
\newcommand{\hst}{{\it HST}}
\newcommand{\cha}{{\it Chandra}}
\newcommand{\teff}{$\mathrm{T_{eff}}$}

\slugcomment{Accepted for publication by {\it The Astrophysical Journal Letters}}

\shorttitle{\hst\ detection of qLMXB companion}
\shortauthors{Edmonds et al.}

\begin{document}

\title{\hst\ Detection of a Quiescent Low Mass X-Ray Binary Companion in
47 Tucanae \altaffilmark{1}}

\altaffiltext{1}{Based on observations with the NASA/ESA {\em Hubble Space
Telescope} obtained at STScI, which is operated by AURA, Inc. under NASA
contract NAS 5-26555.}

\author{Peter D. Edmonds\altaffilmark{2},Craig O. Heinke\altaffilmark{2},
Jonathan E. Grindlay\altaffilmark{2} and Ronald
L. Gilliland\altaffilmark{3}} \altaffiltext{2}{Harvard-Smithsonian Center
for Astrophysics, 60 Garden St, Cambridge, MA 02138;
pedmonds@cfa.harvard.edu; cheinke@cfa.harvard.edu, josh@cfa.harvard.edu}
\altaffiltext{3}{Space Telescope Science Institute, 3700 San Martin Drive,
Baltimore, MD 21218; gillil@stsci.edu}

\begin{abstract}

We present the results of a search for optical counterparts to the two
quiescent low mass X-ray binaries (X5 and X7) in the globular cluster 47
Tucanae, using high quality \cha\ and \hst\ images.  A faint blue
($V=21.7$; $U-V=0.9$) star within 0\farcs03 of the eclipsing system X5
shows variability on both short and long timescales, and is the counterpart
of the X-ray source. The colors and variability of this object are
consistent with the combination of light from an accretion disk and a red
main sequence star (possibly somewhat larger than a normal MS star with
similar luminosity). No evidence is found for a star showing either
variability or unusual colors near the position of X7, but a probable
chance superposition of a star with $V=20.25$ limits the depth of our
search.

\end{abstract}


\section{Introduction}

It has long been thought that quiescent low mass X-ray binaries (qLMXBs)
dominate the most luminous of the dim sources in globular clusters (Hertz
\& Grindlay 1983, Verbunt et al. 1984). Recent observations using
Chandra/ACIS imaging and spectroscopy have demonstrated that this is indeed
the case. Two systems, X5 and X7 in the massive globular cluster 47 Tuc
were previously suspected to be qLMXBs (Hasinger, Johnston \& Verbunt
1994, Verbunt \& Hasinger 1998) but the sensitivity and resolution of
\cha\ was required to confirm this suspicion (Grindlay et al. 2001a,
hereafter GHE01a and Heinke et al. 2001a, in preparation, hereafter
HGL01). One qLMXB has also been found in each of NGC 6397 (Grindlay et
al. 2001b; hereafter GHE01b) and $\omega$ Cen (Rutledge et al. 2001).
These 4 qLMXB systems all have thermal spectra that are well modeled by
hydrogen atmospheres of hot neutron stars (NSs), with no power law
components required. None of them are obviously variable with the exception
of X5 (HGL01) which shows deep eclipses as well as dips showing increased
neutral hydrogen (X7 shows marginal evidence for a 5.5 hr period).

The logical extension of this work is to search for optical counterparts to
these sources, using the potent combination of \cha\ and \hst.  With
astrometric errors $<$ 0\farcs1 routinely being achieved for X-ray sources,
optical identifications are being reported with unprecedented
frequency. These identifications include cataclysmic variables (CVs) and BY
Draconis variables (GHE01a, GHE01b), millisecond pulsars (MSPs; Edmonds et
al. 2001, hereafter EGH01 and Ferraro et al. 2001) and active LMXBs (Heinke
et al. 2001b and White \& Angelini 2001).

Searches for optical counterparts to qLMXBs have been less successful. No
counterpart has been found for the NGC 6397 qLMXB (any possible companion
has \mv > 11; GHE01b) while the $\omega$ Cen qLMXB lies outside the field of
view (FOV) of current \hst\ datasets and stellar crowding will limit deep
searches from the ground.  Here, we report the use of high quality \cha\
and \hst\ data to search for optical counterparts to the 47 Tuc qLMXBs X5
and X7.  We have discovered a faint, blue and variable counterpart to the
eclipsing X5, as reported briefly in GHE01a. This detection, combined with
the well determined period, distance, inclination and X-ray spectrum of X5
makes this the best constrained qLMXB known.  We also report limits on the
qLMXB X7.  The astrometry, photometry and time series for both of these
searches are described below.

\section{Observations and Analysis}

Details of the \cha\ data used here are given by GHE01a and HGL01.  The
qLMXBs X5 and X7 have 4576 and 5488 counts respectively (over the 72 ksec
observation), with internal, 1$\sigma$ errors of 0\farcs0082 and
0\farcs0089 respectively.  To search for optical counterparts to X5 and X7,
two \hst\ datasets have been analysed, the 8.3 d observations of Gilliland
et al. 2000 (GO-8267: July 3 1999 to July 11 1999) and the archival data
of Meylan obtained in three different epochs with $\sim$ 2 year spacings
(GO-5912: October 25 1995; GO-6467: November 3 1997; GO-7503: October 28
1999). The Gilliland data provides exquisite $V$ and $I$ time series (with
some $U$ data) and the Meylan data provides F300W images in the first two
epochs and F300W and F555W images in the third epoch (with limited time
series information in each epoch).

\subsection{Astrometry}

Using the zeropoint positional offsets between the \cha\ and \hst\
coordinate frames, the region within $\sim$2\farcs0 of the nominal X5
position lies outside the FOV of the Gilliland data set, but is found on
the inner part (with respect to cluster center) of the WF4 chip in the
Meylan data. Since no \cha\ source in the WF4 FOV with $>50$ counts (not
including X5) currently has a plausible optical counterpart, we used the PC
astrometry to align the X-ray and optical coordinate frames (incurring a
systematic chip-to-chip error, assumed to be 0\farcs05, which dominates the
total error budget). After this correction we found that only three stars
are within 0\farcs5 of the nominal X5 position, with separations of
0\farcs033 (0.6$\sigma$; C1), 0\farcs23 (4.5$\sigma$; C2) and 0\farcs292
(5.7$\sigma$; C3). The finding chart shows the F300W (Fig. \ref{fig1}a) and
F555W (Fig. \ref{fig1}b) images, from Meylan epoch 3, for the region around
X5 and the insets in Fig. 1a show epochs 1 (`U(1)') and 2 (`U(2)').  Since
faint red MS stars appear brighter in the F555W image than in the F300W
image, Figures 1a and 1b show that C1 has a blue color. However, C3 (just
outside the 5$\sigma$ error circle to the NE) has an even stronger blue
color and so is also a potentially viable optical counterpart if the
astrometric shift between the PC and WF4 chips is much larger than
assumed. This ambiguity is resolved by noting that C1, unlike C3, is
clearly brighter in epoch 1 than in epoch 2 (see inset) confirming it as
the optical counterpart (hereafter \xopt).

%

\begin{figurehere}
\vspace*{0.5cm}
\hspace*{-0.2cm}
\epsfig{file=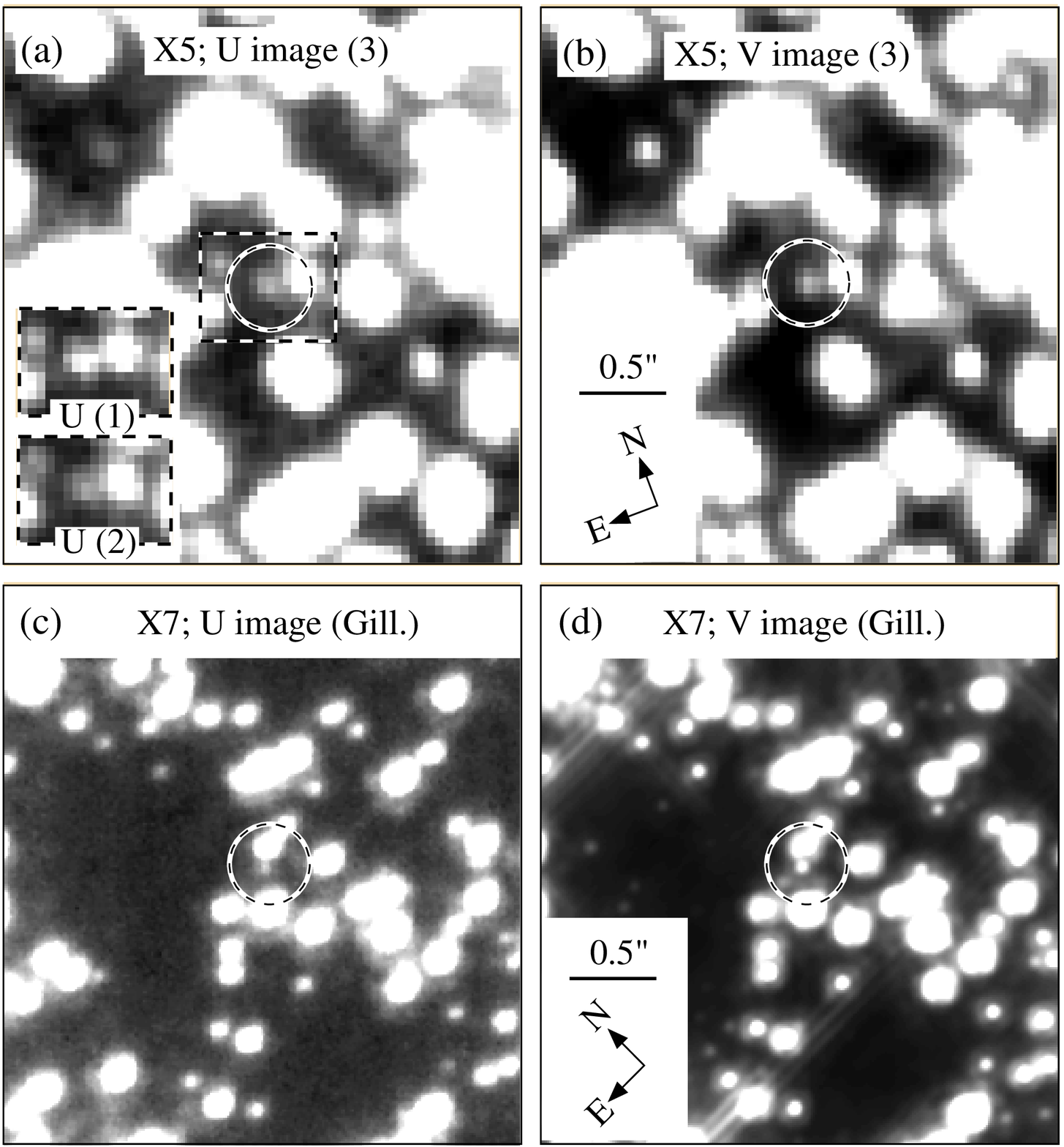,width=8.5cm}
\vspace*{0.2cm}
\caption{Finding charts for X5 and X7. For X5 the 3rd epoch combined F300W
(`U'; Fig. \ref{fig1}a) and F555W (`V'; Fig. \ref{fig1}b) images are
shown. The X5 companion (\xopt) is near the center of the 5$\sigma$ error
circle, and the inset shows the area around \xopt\ from the 1st and 2nd
epochs (note the clear variability of \xopt).  For X7 the deep, oversampled
combined F336W (`U'; Fig. \ref{fig1}c) and F555W (`V'; Fig. \ref{fig1}d)
images from Gilliland et al. (2000) are shown.  The star N1 is found near
the center of the 20$\sigma$ error circle.}
\vspace*{0.5cm}
\label{fig1}
\end{figurehere}

The region within a few arcsec of X7 falls on the PC images of both the
Gilliland and Meylan data sets. Since 17 \cha\ sources have likely optical
counterparts on the Gilliland PC image, we have corrected for small linear
terms in the residual astrometric errors between \cha\ and \hst\ using
least squares fitting. We computed the positional errors for X7 by adding
the systematic errors to the random {\tt wavdetect} errors in quadrature,
resulting in 1$\sigma$ errors of 0\farcs0065 in RA and 0\farcs0088 in Dec
(see Fig. \ref{fig1}c and \ref{fig1}d, where 20$\sigma$ error circles are
shown). The three nearest stars to X7 in the Gilliland \hst\ image are
0\farcs019 (2.3$\sigma$), 0\farcs12 (14.7$\sigma$) and 0\farcs23
(28.6$\sigma$) away (N1, N2 and N3 respectively). Clearly, astrometrically,
only N1 ($V=20.25$; $U-V=1.72$; \mv\ = 6.8) is a viable candidate for the
optical companion of X7. Given the FOV of the PC and the detected number of
stars on the PC chip with $V<20.25$ (6367), only $6.3\times10^{-3}$ stars
are expected within 0\farcs019 of N1, assuming constant density over the PC
FOV.

\subsection{Photometry}

The Gilliland dataset photometry (containing only X7) is described in
Gilliland et al. (2000) and Albrow et al. (2001). The photometry for the
Meylan observations (containing X5 and X7) was based on combining the
images at each epoch using drizzle routines (Hook, Pirzkal, \& Fruchter
1999) in {\tt STSDAS} and then using PSF-fitting in {\tt DAOPHOT} to
calculate instrumental magnitudes. The F555W filter is a good approximation
to Johnson $V$ (Holtzman et al. 1995), but F300W differs significantly from
the nearest Johnson filter ($U$). Therefore, we used ground-based
photometry of 47 Tuc (Sills et al. 2000) and matching of main sequence (MS)
turnoffs between the \hst\ and ground-based datasets to calculate the
zeropoint and then applied corrections to F300W-$V$ (by measuring MS
ridgelines) to convert it to $U-V$. By definition this MS-ridgeline
technique is only applied to stars (like \xopt) with colors ranging from
the main sequence turn-off to the detected end of the MS.

\begin{figurehere}
\vspace*{-0.1cm}
\hspace*{-0.8cm}
\epsfig{file=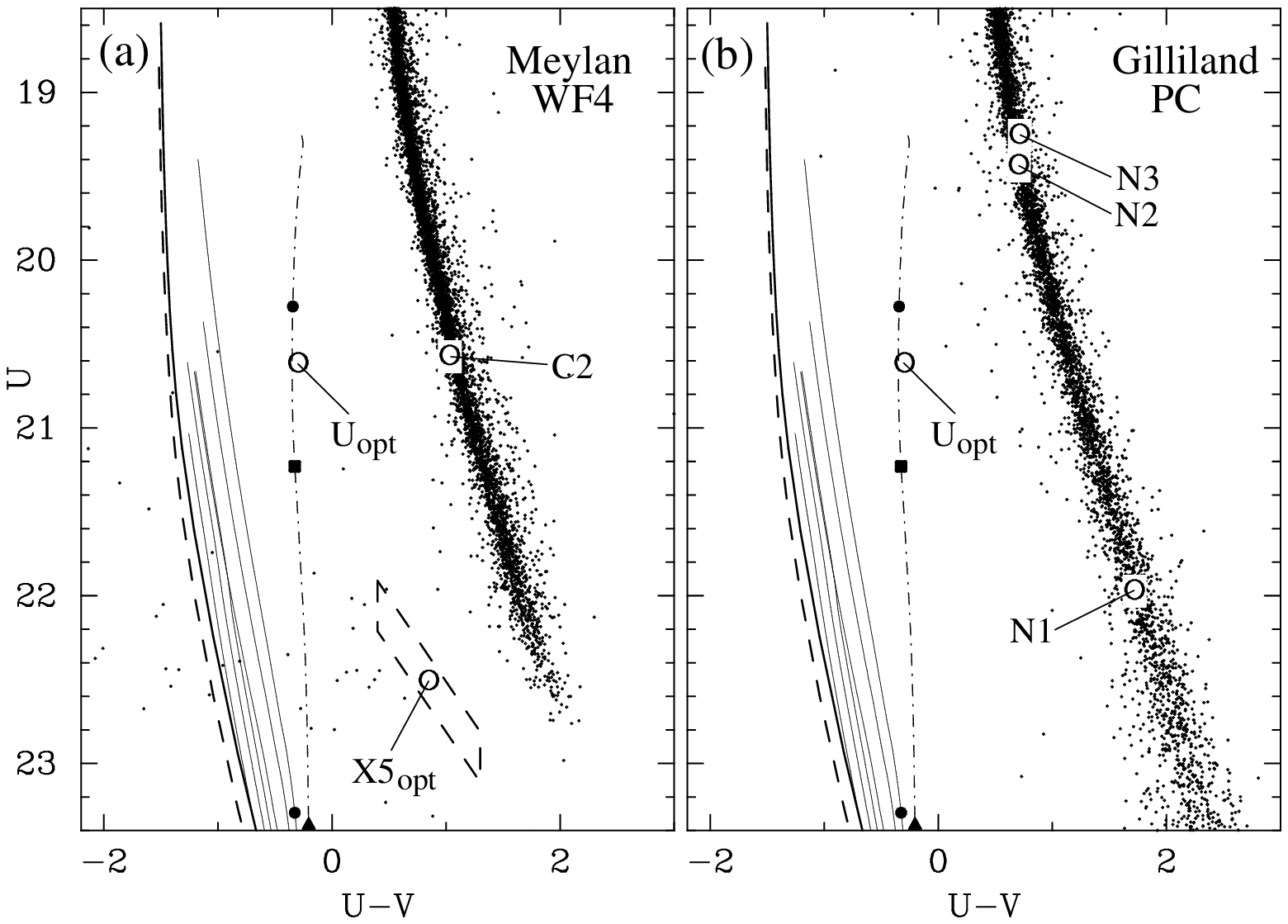,width=9.5cm}
\vspace*{-0.2cm}
\caption{$U$ vs $U-V$ CMDs for the Meylan WF4 chip (containing \xopt;
Fig. \ref{fig2}a) and the Gilliland PC chip (containing X7;
Fig. \ref{fig2}b). For \xopt\ we show ranges in magnitude and color given
limits on the variability. Near neighbors to \xopt\ and X7 are shown (apart
from C3 which is not detected in the F555W image.  Also shown are CO WD
cooling sequences from Bergeron, Wesemael \& Beauchamp (1995; thick lines)
and He WD cooling sequences from Serenelli et al. (2001; thin lines). The
dot-dashed sequence is the lowest mass model (the MSP counterpart \uopt\
from EGH01 is also shown).}
\vspace*{0.5cm}
\label{fig2}
\end{figurehere}

Since this technique is non-standard we have performed two consistency
checks with other calibration methods.  We applied this technique to the
Meylan PC data and performed a star-by-star comparison between our
photometry and the Gilliland et al. (2000) photometry. Mean differences
between the two photometric systems were $< 0.05$ mag in both $U$ and $V$.
A star-by-star comparison between the MS-ridgeline $V$ calibration for the
Meylan WF4 chip (containing \xopt) and the standard calibration of Holtzman
et al. (1995) applied to a 47 Tuc F555W image from the archive (program
GO-6095), also gave mean errors < 0.05 mag.  Combined with the 0.1 mag rms
internal error in $U-V$ at the $U$ mag of \xopt, we estimate absolute
errors for \xopt\ of $\sim$0.2 mag in both $U$ and $V$.

The color magnitude diagrams (CMDs) for the Gilliland PC and Meylan WF4
images are shown in Fig. \ref{fig2}.  Fig. \ref{fig2}a shows the mean epoch
3 CMD position of \xopt\ ($V=21.7$; $U-V=0.9$; \mv\ = 8.2), along with
reasonable ranges in magnitude and color given the variability (see below).
With \mv$=8.2$, \xopt\ has a similar absolute magnitude to that of the
qLMXBs Cen-X4 (\mv$=7.5-8.5$; Chevalier et al. 1989) and Aql X-1
(\mv$=8.1$; Chevalier et al. 1999).  Also shown are CO and He WD cooling
sequences (see caption). Clearly, \xopt\ is unlikely to be either a CO WD
or a He WD, unlike the MSP companion \uopt\ (EGH01).  Instead, it is more
likely that the CMD position of \xopt\ represents the sum of a red MS star
and a blue component from an accretion disk (see below).

\begin{figurehere}
\vspace*{-1.0cm}
\hspace*{-0.3cm}
\epsfig{file=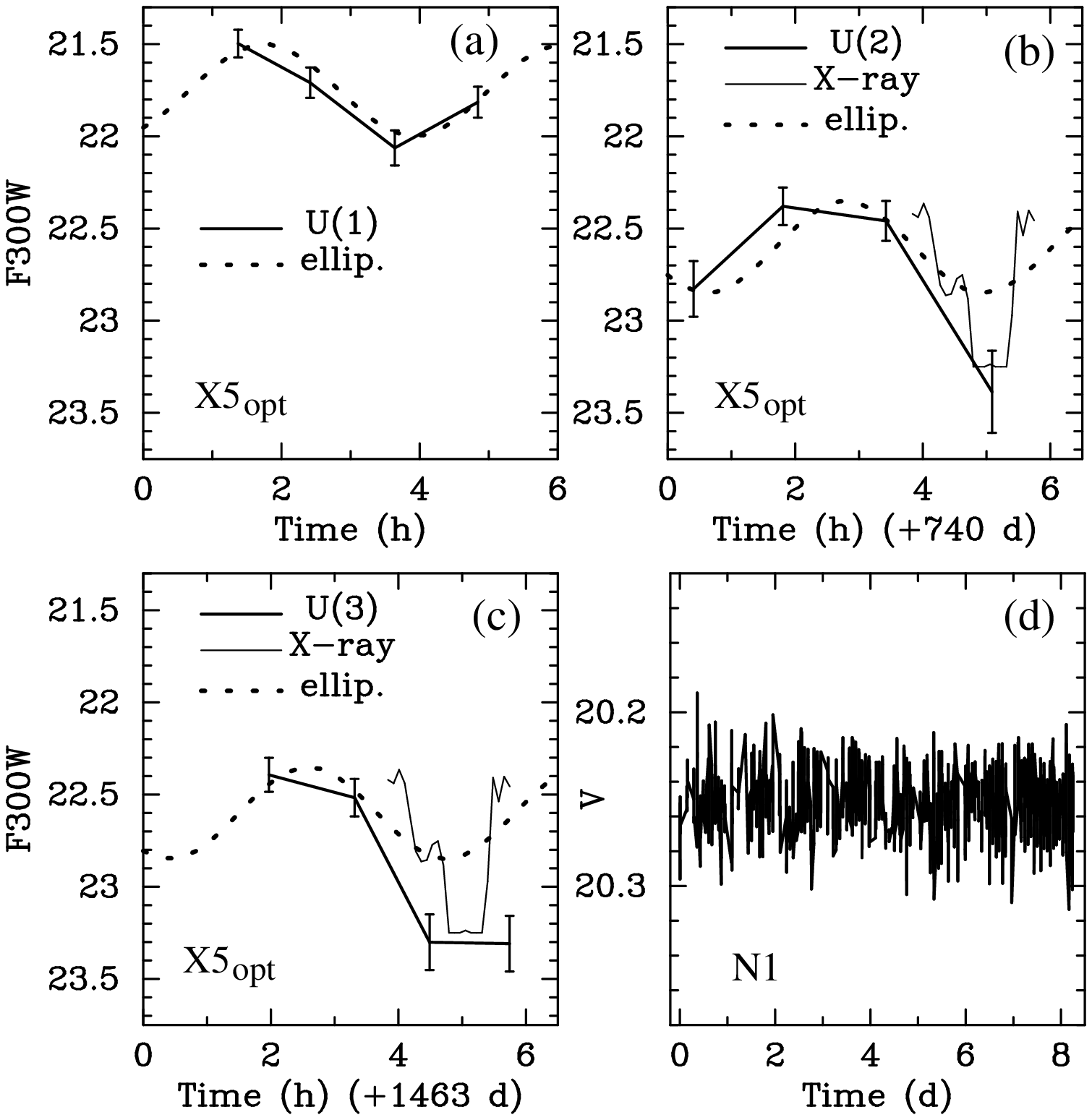,width=9.2cm}
\vspace*{-2.2cm}
\caption{F300W time series for \xopt\ for epochs 1, 2 and 3 (Figures
\ref{fig3}a, b and c respectively) with time offsets between epochs as
shown. For \xopt\ a simple ellipsoidal model (`ellip.'), eclipse not
included, is shown (zeropoints adjusted to the data).  Also shown is
the eclipsed portion of the X-ray phase plot (units converted into time)
from HGL01.  The 1$\sigma$ error bars are the median time series rms
for stars within 0.25 mag of a given magnitude level. Figure \ref{fig3}d
shows the $V$-band time series for N1, the nearest possible counterpart to
X7.}
\vspace*{0.5cm}
\label{fig3}
\end{figurehere}

Fig. \ref{fig2}b shows the Gilliland CMD for X7 showing the position of the
three nearest stars (N1, N2 and N3). The only viable counterpart
astrometrically, N1, falls very close to the MS ridge line (also in $V$ vs
$V-I$) and therefore appears like a normal MS star (unlike \xopt). This CMD
position is consistent, within the errors, with the position in the Meylan
data, and the F300W magnitudes from the three epochs are consistent with
non-variability, again unlike \xopt. This suggests that N1 is probably not
the X7 counterpart, despite the good astrometric match.  Assuming that the
real counterpart falls in the less confused part of a 5-$\sigma$ error
circle we set limits on its detection of $U >$ 23, $V >$ 23, $I >$ 22
(using the Gilliland data).

\subsection{Time Series}
\label{timeser}

Since \xopt\ is near (or beyond) the limit of detectability in individual
F300W exposures, the time series for \xopt\ were calculated by co-adding
groups of 3-4 images. Figures \ref{fig3}a, b and c show the F300W time
series for the 3 different epochs. Also shown are the eclipsed portion of
the X-ray phase plot from HGL01 (units converted into time) and a 4.333
hour period sinusoid, as appropriate for X5 but with the 8.666 hour X-ray
period divided by two to simulate a double-peaked (ellipsoidal) time
series. Eclipses are not included in this model. This model has been
shifted in time and magnitude so that it plausibly matches the data for
each epoch (the period is not known with sufficient accuracy to phase
correct from \cha\ to different \hst\ epochs).  Significant variability is
seen within all three epochs and \xopt\ is clearly brighter in epoch 1 than
in the 2nd and 3rd epochs (see Fig \ref{fig1}a). This longterm variability
is further evidence for the presence of an accretion disk.  Note the deep
eclipse observed in both epochs 2 and 3. The F300W eclipse appears to be
significantly wider than in X-rays, though we do not observe the system
coming out of eclipse. The turnover at the beginning of the second epoch
(near Time$=0$ hr) may represent a minimum from ellipsoidal variability,
since the timescale of variability agrees well with the ellipsoidal
model. The 1st epoch observations may also represent ellipsoidal variations
rather than an eclipse, since in its brighter state the relative brightness
of the disk compared to the secondary should be enhanced and the eclipse
depth should increase.   The $V$-band variations, not plotted here, show neither
an eclipse nor clear evidence for ellipsoidal variations (expected
to have a smaller amplitude than in F300W).

No suggestion of variability is present in the N1 time series
(Fig. \ref{fig3}d) and no significant signal is seen in the N1 power
spectrum ($V$ or $I$), including the possible 5.5 hr period noted by
HGL01. The highest peak in the $V$-band corresponds to a period of 2.52
hours (or twice this), with a false-alarm probability = 0.27.  The
corresponding $V$ amplitude is 0.0043 $\pm$ 0.0011 ($<$ 4-$\sigma$,
insignificant for a blind search; similar results hold for $I$).  If N1
{\it is} the X7 counterpart and is close to filling its Roche lobe then an
inclination $< 2.5$\degr\ is required to reduce the amplitude for
ellipsoidal variations from the maximum expected value of $\sim$0.1, for
90\degr\ inclination, to $<0.0043$.  This implies that N1 is unlikely to be
the X7 companion, and we have calculated the brightness limit for a faint
variable star (lying near the line of sight of N1) to be missed by our
variability search.  The X7 coordinates are so close to N1 that it will be
included in any time series extraction.  A star at $V=22.9$ with intrinsic
variations of 0.1 mag superimposed on the time series of N1 would yield an
8-$\sigma$ detection (versus the highest detected peak at $\sim$
4-$\sigma$). This time series limit of $V\sim23$ for a companion to X7 will
decrease for inclinations $< 90$\degr.

\section{Discussion}

Using the stellar models of Bergbusch \& Vandenberg (1992), we estimated
the brightest possible secondary consistent with our photometry (\teff$ =
4100$K, $V=21.7$ and mass = 0.53\mdot).  Using the secondary radius, the
X-ray luminosity of X5 and the binary separation (from Kepler's Third Law)
we estimate that the maximum luminosity from heating of the secondary by
the NS (when measured as a fraction of the secondary luminosity) is 2.7\%.
Therefore, secondary heating probably makes only a small contribution to
the variability described above. The dominant sources of short-term
variability are likely to be a combination of eclipses of the disk and hot
spot by the MS star, ellipsoidal variations and flickering. Further
observations are required to better define this variability. 

The likely presence of an accretion disk in \xopt, from variability and the
blue color, may appear to be inconsistent with the lack of X-ray evidence
for accretion currently in the X5 system, which should yield either
long-term X-ray variations or a power law component, neither of which are
seen (HGL01).  One possible explanation is that the X5 secondary is no
longer filling its Roche lobe causing it to be detached from the disk. Such
a disk would no longer be accreting matter from the secondary, possibly
causing it to enter a long-term quiescent phase with low density and little
or no accretion onto the NS. The X5 disk does appear to be relatively faint
compared to 47 Tuc CVs, since the $U-V$ color of \xopt\ (0.9) is much
redder than that of the 47 Tuc CVs V1, V2, W1 and W2 with $U-V$ colors
ranging from $-1.25$ to $-0.4$ (Edmonds et al. 2001, in preparation).


To test this `detached disk' theory we have estimated the degree to which
\xopt\ fills its Roche-lobe as defined by $F$, the ratio between the
stellar radius and the Roche lobe radius.  Using the Roche-lobe formula
from \citet{pac71} ($r/a=\mathrm{0.462[(M_{opt}/(M_{NS}+M_{opt})]^{1/3}}$,
where $r$ is the Roche-lobe radius, $a$ is the binary separation,
$\mathrm{M_{opt}}$ is the mass of \xopt, and $\mathrm{M_{NS}=1.4}$\mdot),
the stellar radius for a 4100 K model and the binary separation, we find
that \xopt\ has $F=0.6$, underfilling its Roche lobe. Fainter cooler
secondaries will underfill their Roche lobes by slightly larger amounts
(e.g. a star with \mv\ = 10.0 has $F=0.5$). This behavior is consistent
with the `detached disk' theory given above, but would be inconsistent with
the possible detection of ellipsoidal variations of relatively large
amplitude, requiring the secondary to have $F\sim$1.0. The latter
possibility would suggest that the X5 secondary is either bloated or
slightly evolved, as appears to be the case for some of the CVs in NGC 6397
(Grindlay et al. 2001, in preparation) and as might be expected for a star
undergoing mass loss.

If the 5.5 hr X-ray period for X7 (HGL01) is real, and if the X7 secondary
underfills its Roche lobe by about the same amount as \xopt, then
\mv$\sim10.6$ and $V\sim24.1$, beyond our variability detection limits and
beyond our CMD search limit except with the presence of a reasonably bright
disk.  The close proximity of N1 only $\sim$0\farcs02 from the line of
sight of X7, clearly makes prospects for such an optical identification
difficult. Alternatively, the period could be longer and N1 could be the
counterpart, however the complete lack of evidence for a disk or
variability rules against this possibility.

The logical follow-up to these observations are spectroscopic studies to
measure the radial velocity amplitude of absorption lines from \xopt.
This, combined with the known inclination and spectroscopic and photometric
determinations of $\mathrm{M_{opt}}$ would give an estimate of the mass of
the X5 NS.  Using the X-ray spectrum constraints on the NS radius and
redshift (HGL01) combined with the NS mass would give the first compelling
test of the equation of state of a NS. Also of interest would be the
detection of emission lines in the optical spectrum.  Since there is
evidence for an accretion disk (from this work) and hot gas in the system
(from the X-ray light curve), we expect strong disk or coronal emission
lines to be superimposed on the absorption line spectrum of the
secondary. Study of the emission line profiles can test for mass outflow
(visible as P Cygni profiles) from the system, or detect evidence for a
bipolar jet (visible as broadened emission lines).


\acknowledgments

We thank Ata Sarajedini, Raja Guhathakurta, and Justin Howell for
contributing to the photometric analysis and Bryan Gaensler and Frank
Verbunt for helpful comments on the manuscript.  This work was supported in
part by STScI grants GO-8267.01-97A (PDE and RLG) and HST-AR-09199.01-A
(PDE).




\begin{deluxetable}{cllrrccr}
\tabletypesize{\footnotesize}
\tablecolumns{8}
\tablewidth{0pc}
\tablecaption{Positional, photometric and time series information for \xopt}
\tablehead{ \colhead{qLMXB} &
 \colhead{RA\tablenotemark{a}} & \colhead{Dec\tablenotemark{a}} &
 \colhead{X\tablenotemark{b}} & \colhead{Y\tablenotemark{b}} &
 \colhead{$U$\tablenotemark{c}} & \colhead{$V$\tablenotemark{c}} & 
 \colhead{period\tablenotemark{d} } \\
\colhead{} & \colhead{(J2000)} & \colhead{(J2000)} & \colhead{} &
\colhead{} &  \colhead{} & \colhead{} &
\colhead{(hours)} 
}
\startdata

X5 & 00 24 00.991(1) & $-$72 04 53.202(7) & 60.7  & 263.0 & 22.5(2) & 
  21.7(2)  & 8.67 \\
X7 & 00 24 03.528(1) & $-$72 04 51.938(6) & 498.5 & 775.5 & \nodata & 
  \nodata  & 5.50 \\

\tablenotetext{a}{Coordinates in MSP astrometric system; Freire et al. (2001)}
\tablenotetext{b}{Pixel coordinates of X-ray sources using the {\tt STSDAS}
task {\tt METRIC} applied to archival images u2ty0201t (X5) and u5jm070cr (X7)}
\tablenotetext{c}{Epoch 3 mean magnitudes for \xopt}
\tablenotetext{d}{Periods from HGL01; value for X7 is from a
marginal detection of variability}


\enddata
\end{deluxetable}



\end{document}